%% file: SO-JJ-half-vortex_arxiv-submission.tex
\documentclass[twocolumn,english,aps,prl,floatfix,amssymb,superscriptaddress,longbibliography]{revtex4-1}
\usepackage{color}
\definecolor{note_fontcolor}{rgb}{0.800781, 0.0, 0.0}
\usepackage{babel}
\usepackage{braket}
\usepackage{amsmath}
\usepackage{amssymb}
\usepackage{graphicx}
\usepackage{esint}
\usepackage{bbold}
\usepackage[unicode=true,pdfusetitle,
 bookmarks=true,bookmarksnumbered=false,bookmarksopen=false,
 breaklinks=false,pdfborder={0 0 0},backref=false,colorlinks=true,citecolor=blue]
 {hyperref}

\makeatletter

\@ifundefined{textcolor}{}
{%
 \definecolor{BLACK}{gray}{0}
 \definecolor{WHITE}{gray}{1}
 \definecolor{RED}{rgb}{1,0,0}
 \definecolor{GREEN}{rgb}{0,1,0}
 \definecolor{BLUE}{rgb}{0,0,1}
 \definecolor{CYAN}{cmyk}{1,0,0,0}
 \definecolor{MAGENTA}{cmyk}{0,1,0,0}
 \definecolor{YELLOW}{cmyk}{0,0,1,0}
}

\makeatother

\begin{document}

\title{Fractional Josephson Vortices and Braiding of Majorana Zero Modes in Planar Superconductor-Semiconductor Heterostructures}

\author{Ady Stern}
\affiliation{Department of Condensed Matter Physics, Weizmann Institute of Science, Rehovot, Israel 76100}
\author{Erez Berg}
\affiliation{Department of Condensed Matter Physics, Weizmann Institute of Science, Rehovot, Israel 76100}
\affiliation{Department of Physics and the James Frank Institute, University of Chicago, Chicago, IL 60637, USA}

\begin{abstract}
We consider the one-dimensional (1D) topological superconductor that may form in a planar superconductor-metal-superconductor Josephson junction in which the metal is is subjected to spin orbit coupling and to an in-plane magnetic field. This 1D topological superconductor has been the subject of recent theoretical and experimental attention. We examine the effect of perpendicular magnetic field and a supercurrent driven across the junction on the position and structure of the Majorana zero modes that are associated with the topological superconductor. In particular, we show that under certain conditions the Josephson vortices fractionalize to half-vortices, each carrying half of the superconducting flux quantum and a single Majorana zero mode. Furthemore, we show that the system allows for a current-controlled braiding of Majorana zero modes.
\end{abstract}
\maketitle


{\it Introduction.---}Significant progress has been made in recent years towards realizing topologically-protected zero modes in condensed matter systems~\cite{Lutchyn2018majorana,Aguado2017majorana}. Among their special properties, these states, known as Majorana zero modes (MZMs), attracted a lot of attention because of their non-Abelian exchange properties~\cite{Alicea2012,Beenakker2013}, that may enable them to store and manipulate quantum information in a robust topologically-protected manner.

Majorana zero modes appear, for example, at vortex cores of two-dimensional topological superconductors~\cite{Kopnin1991mutual,Read2000paired} and at the ends of one-dimensional topological superconductors~\cite{Kitaev2001unpaired}. Topological superconductivity can be engineered in carefully designed hybrid systems of conventional superconductors and conventional materials with strong spin-orbit coupling~\cite{Lutchyn2010majorana,Sau2010generic,Oreg2010helical,Nadj2013proposal,Klinovaja2013topological}; several platforms realizing this state of matter have been explored~\cite{Mourik2012signatures, Das2012zero, Albrecht2016exponential, Nadj-Perge2014observation}.
In particular, much effort has been devoted to systems of semiconductor nanowires coupled to superconductors.
There is mounting evidence that the long-sought Majorana zero modes appear at the ends of the wires when the system enters the topological phase.

These remarkable developments call for the next steps towards demonstrating non-Abelian statistics, and raise the question of what is the ideal physical platform to control, manipulate and probe Majorana zero modes. Looking further, one would ultimately like to be able to construct complex networks of many interlinked zero modes and be able to manipulate them at will. To that end, physical realizations of Majorana zero modes that allow for new experimental knobs to control them are highly desirable. Recent experiments in two dimensional electron gases (2DEGs) with strong spin-orbit coupling (SOC) have demonstrated robust proximity coupling to superconductors~\cite{Shabani2016two,Suominen2017zero}, opening a new promising path in these directions.

\begin{figure}[b]
\begin{center}
\includegraphics[width=0.5\textwidth]{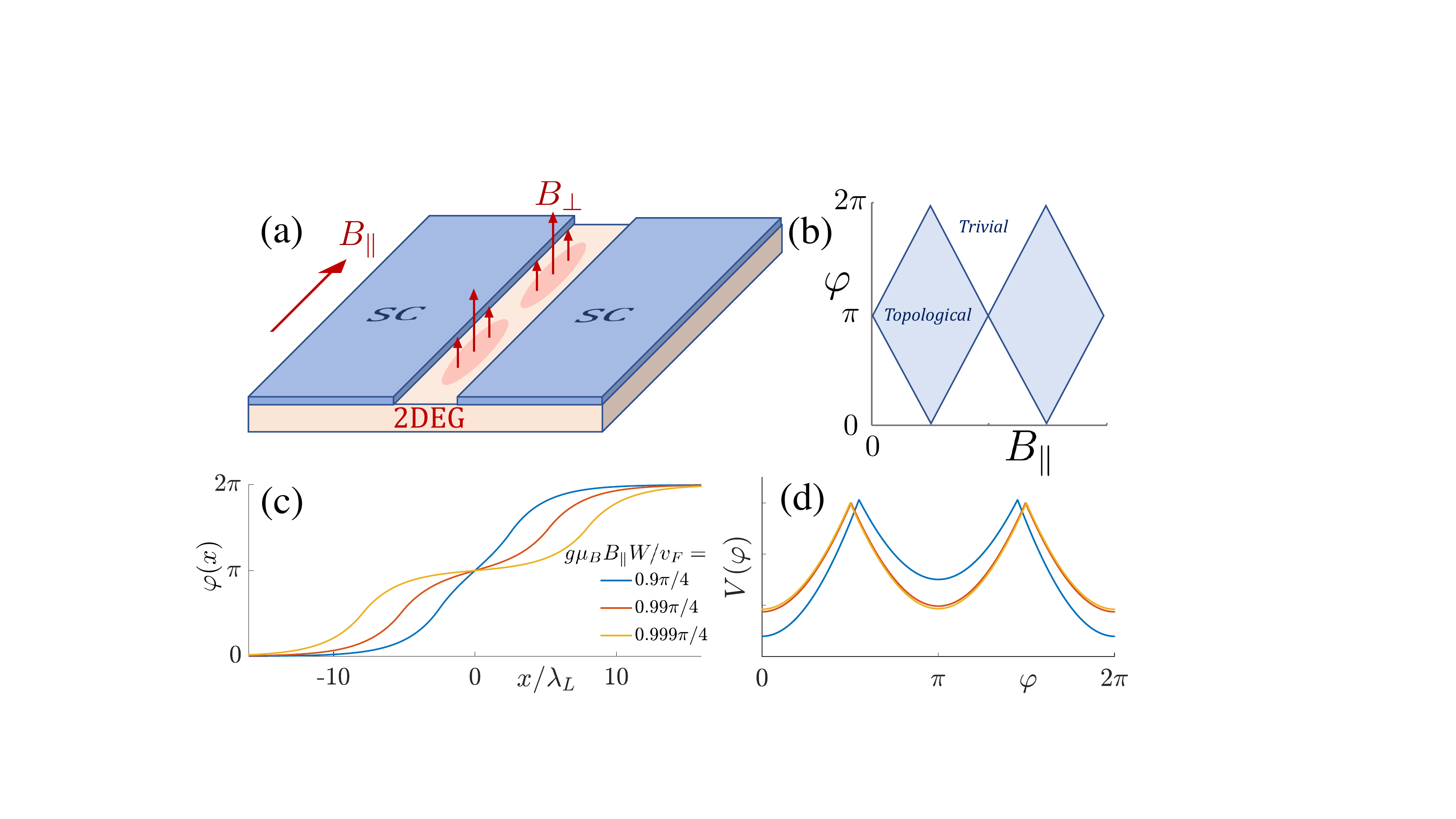}
\caption{(a) Planar Josephson junction connecting two SC leads through a 2DEG with Rashba SOC, in the presence of an in-plane and a perpendicular magnetic field. The Josephson vortices in the junction can split into fractional vortices carrying fractions of the SC flux quantum $hc/2e$. (b) Phase diagram of the junction, viewed as a 1D system, as a function of the phase difference $\varphi$ between the two SCs and the parallel magnetic field (see Ref.~\cite{Pientka2017topological}). (c) $\varphi(x)$ along the junction for different values of $B_\parallel$ close to the critical value, $g\mu_B B_{\parallel,c} W/v_F = \pi/4$, at which the ground state of the junction switches from $\varphi\approx 0$ (within the topologically trivial phase) and $\varphi\approx \pi$ (in the topological phase). (d) Energy-phase relations for the values of $B_\parallel$ in panel (c).}
\label{fig:system}
\end{center}
\end{figure}

In this work, we focus on a new setup proposed recently to realize one-dimensional topological superconductivity in a planar Josephson junction~\cite{Pientka2017topological,Hell2017two,Hell2017Coupling}. The setup is shown schematically in Fig.~[\ref{fig:system}]. Two superconducting films are deposited on top of a 2DEG with Rashba Spin-Orbit coupling. The resulting sub-gap Andreev bound states form an effective one-dimensional system, that can be controlled either by applying a magnetic field parallel to the junction $B_{||}$, or by varying the phase difference $\varphi$ between the two superconductors. It was recently shown~\cite{Pientka2017topological,Hell2017two} that this system can readily enter a one-dimensional superconducting phase with Majorana zero modes at its ends. In particular, for a given magnetic field $B_{||}$ there is a range of phase differences $\varphi_-(B_{||})<\varphi<\varphi_+(B_{||})$ in which the junction hosts such Majorana zero modes at its ends. When the phase difference is $\varphi = \pi$, the topological phase is most stable. Moreover, when the phase difference is not externally controlled and a magnetic field is applied parallel to the junction, the phase can self-tune to a value near $\pi$ via a first-order phase transition, driving the system into the topological phase. Disorder can have a stabilizing effect on the MZMs associated with the topological phase~\cite{Haim2018double}. The studies were motivated by an experiment~\cite{Hart2017controlled} that realized such a system, and were followed by experimental studies that  observed signatures of MZMs in these systems~\cite{Ren2018topological,Fornieri2018evidence}.

Here we add two additional tuning knobs to the set-up, a magnetic field $B_\perp$ perpendicular to the junction and a supercurrent $I$ driven across the junction. We explore the way by which these two knobs may serve to control and manipulate topologically protected zero modes in this system, exploiting its unique properties.

In our discussion we distinguish between cases where screening currents are significant or insignificant. In the first case, realized in junctions that are longer than the Josephson screening length ($L\gg\lambda_J$), the perpendicular magnetic field creates Josephson vortices~\cite{Tinkham2004introduction}. We analyze the structure of these vortices in the junction and find that, if the system is tuned near the $0$ to $\pi$ first-order transition, a Josephson vortex tends to spontaneously ``fractionalize'' into two half-vortices (carrying a flux of $\Phi_0/2 = hc/4e$ each). Each half vortex is effectively a domain wall between the topological and the trivial phases of the junction; as a result, it carries a protected Majorana zero mode. We also analyze the position of MZMs in different geometries and the way it is affected by the screening currents.

For the case where screening currents are insignificant ($L\ll\lambda_J$) we propose a tri-junction structure~\cite{Alicea2011non,Clarke2011majorana} where supercurrents between the different parts of the junction serve to control the location of the Majorana zero modes and their coupling. This control allows for a scheme that braids the Majorana zero modes, thus revealing their non-Abelian properties.

{\it Phase configuration and position of MZMs as a function of $B_\perp$ and $I$.---}We start by considering the effect of $B_\perp$ and $I$ on the phase configuration in the junction. Generally, the phase configuration $\varphi(x)$ in a long Josephson junction is determined by balancing two energies: the magnetic energy, whose density is proportional to $(\partial_x\varphi)^2$, and the washboard potential Josephson energy, whose density depends on $\varphi(x)$ itself through the Josephson energy per unit length, $V\big(\varphi(x)\big)$. The balance leads to the equation~\cite{Tinkham2004introduction}
\begin{equation}
\kappa \partial_x^2\varphi(x)=V'\big(\varphi(x)\big).
\label{current-configuraion}
\end{equation}
Here, $\kappa \equiv \frac{\Phi_0^2}{2\pi \mu_0(w+2\lambda_L)}$, where $w$ is the width of the junction, $\lambda_L$ is the London penetration depth of the superconductor, and $\mu_0$ is the permeability of the vacuum. The current $I$ through the junction constrains the phase to satisfy
\begin{equation}
\frac{2\pi}{\Phi_0}\int_0^LV'\left(\varphi(x)\right)dx =I.
\label{phaseconf}
\end{equation}
The unique properties of the junction we consider are reflected in the potential $V(\varphi)$, as we review below.

When the Josephson coupling $V(\varphi)$ is small (a condition defined more precisely below), Eq.~(\ref{current-configuraion}) may be solved by iterations. At the lowest iteration the right-hand side is set to zero and the phase configuration obtained is
\begin{equation}
\varphi(x)=\varphi_0+\beta x.
\label{no-screening}
\end{equation}
The  magnetic field controls the gradient of the phase and sets $\beta=(2\lambda_L+w)B_\perp/\Phi_0$. The determination of $\varphi_0$, the value of the phase difference at $x=0$, depends on geometry. When the current $I$ is controlled, $\varphi_0$ is found by substituting  (\ref{no-screening}) in the expression for the current across the junction and solving Eq.~(\ref{phaseconf}).
 In contrast, in a flux-loop geometry the current across the junction is not controlled. Rather, it is the flux in the loop that determines $\varphi_0$.

For a given phase configuration, the junction we consider may host MZMs at its ends or at its bulk. An MZM occurs at the $x=0,L$ ends when the phase at these points is within the topological regime, i.e., $\varphi_-+2\pi n<\varphi(x=0,L)<\varphi_++2\pi n$, where $\varphi_\pm$ are the critical values of the phase where the topological transition occurs, that depend on $B_\parallel$ (see Fig.~\ref{fig:system}), and $n$ is an integer.
In contrast, MZMs at the bulk  would occur at the transition points $x_\pm$ between topological and trivial segments, i.e., points defined by
$\varphi(x_\pm)=\varphi_\pm$.

The spatial extent of the MZMs is determined by the energy gap. For the MZMs at the ends of the junction the gap is the bulk gap $\Delta_0$ and the localization length is $\xi_M^0=hv/\Delta_0$, with $v$ a characteristic velocity. For the MZMs {at a domain wall between the two phases in the bulk} of the junction the gap vanishes at the critical points $\varphi_\pm$ and is proportional to $|\varphi-\varphi_\pm|$ close to those points. The phase varies linearly with position close to $x_\pm$. As long as this variation is slow, we may define a local gap $\Delta(x)\approx \Delta_0\frac{|x-x_\pm|}{\Delta x}$, where {$\Delta x \sim 1/\beta \propto 1/B_\perp$} is the distance over which the phase varies from $\varphi_\pm$ to the value where the gap is maximal.
With the gap varying linearly with position, we estimate $\xi_M$ by solving $\xi_M=\hbar v/(\Delta \xi_M)$, which gives $\xi_M=\max{\left (\xi_M^0,\sqrt{\xi_M^0\Delta x}\right )}$. For small $B_\perp$, the MZMs are well separated and their coupling is small, independently of the ratio of $\Delta x$ to $\xi_M^0$. This is since the distance between the MZMs scales with $1/B_\perp$, while $\xi_M$ scales at most as $1/\sqrt{B_\perp}$.

As an example to the way MZMs may be manipulated we consider a junction that hosts four MZMs. That may happen when the phase $\varphi(x)$ is in the topological regime at both ends $x=0,L$, but with values of $n$ that differ by one.  Two of these modes $\gamma_{1,4}$ are located at the junction ends $x=0,L$. The other two, $\gamma_{2,3}$ are located at the points $x_{\pm}$ where $\varphi=\varphi_{\pm}$, respectively. Now, when a current is applied, the phase configuration shifts according to (\ref{phaseconf}). For weak currents, the zero modes at $x=0,L$ do not move, but the points $x_{\pm}$ move, keeping $x_+ -x_-$ constant. Thus, the coupling between $\gamma_1$ and $\gamma_2$ and the coupling between $\gamma_3$ and $\gamma_4$ would be affected by a current driven through the junction. A small variation of the perpendicular magnetic field, on the other hand, would affect all distances between the zero modes, and therefore all nearest-neighbors couplings.

Overall then, in the limit of weak Josephson coupling, well separated MZMs may be created in pairs, moved, and annihilated in pairs by varying $\varphi_0$ and $\beta$, i.e., by varying $B_\perp$ and $I$. For a fixed $B_{||}$ MZM pairs are created and annihilated at the ends. Below, we will analyze the way $B_\perp$ and $I$ may be employed to braid pairs of MZMs. Before doing so, however, we turn to examine what happens when the Josephson coupling is not weak.

For the context of the present discussion the strength of the coupling is determined by the ratio of the Josephson screening length $\lambda_J$ to the junction length $L$. Our discussion has so far assumed $L\ll\lambda_J$, a case in which the magnetic field created by the Josephson current is negligible, and Eq. (\ref{no-screening}) is a good approximation. In the opposite limit, $L\gg\lambda_J$, the magnetic field created by the Josphson currents is not negligible, the flux is either screened to be within a distance $\lambda_J$ from the junction's ends or penetrates the junction in the form of Josephson vortices, and the phase configuration is more complicated than the form (\ref{no-screening}).

For an SIS junction, where $V(\varphi)=\epsilon_J(1-\cos{\varphi})$, this limit is well studied. Eq. (\ref{current-configuraion}) becomes the Sine-Gordon equation, and the Josephson vortex is a soliton that connects minima points of $\varphi_n=2\pi n$~\cite{Tinkham2004introduction}. When no vortex penetrates the junction, the phase generally varies only over a distance $\lambda_J$ from one of the junction's ends, and assumes one of the values $\varphi_n$ further into the junction's bulk. The end where the phase varies is the end where the Josephson currents flows, and it depends on the geometry (see Supplementary Material). When vortices penetrate the junction, they connect between neighboring values of $\varphi_n$.

\begin{figure*}[t]
\begin{center}
\includegraphics[width=1.0\textwidth]{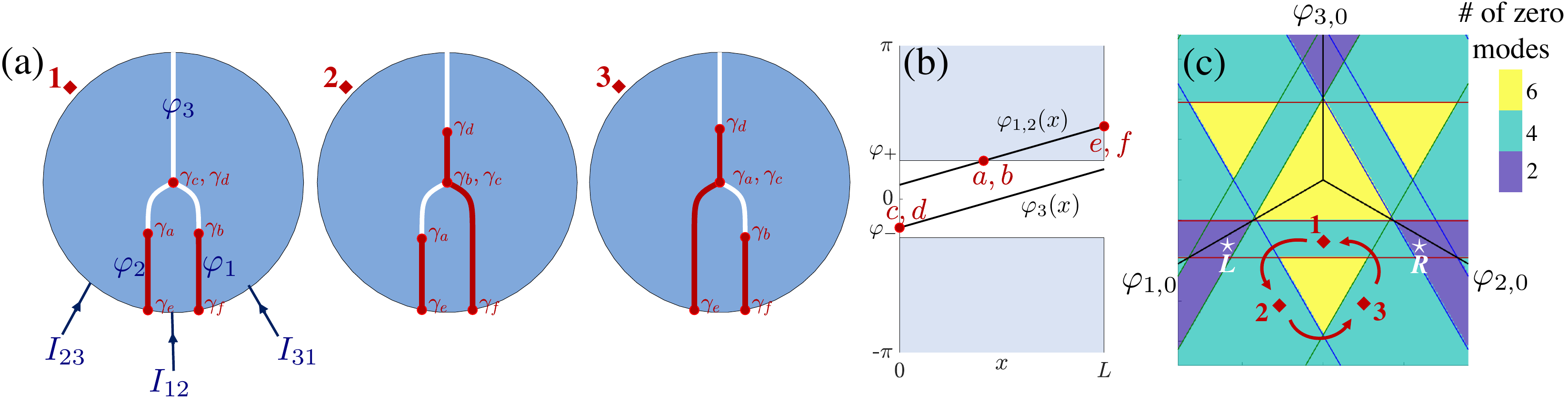}
\caption{(a.1--3) Proposed setup for braiding MZMs. Three Josephson junctions meet at a T junction. The perpendicular magnetic field makes the superconducting phases $\varphi_{1,2,3}$ vary along the junction. Whenever $\varphi_{1,2,3}(x)$ crosses one of the critical values $\varphi_{\pm}$ or their $2\pi$ equivalents, the corresponding junction undergos a topological phase transition. Red (white) lines indicate segments of the junction that are in the topological (trivial) phase, respectively; the MZMs $\gamma_{a-f}$ that form at the boundaries of the topological regions are indicated by red circles. The externally injected currents $I_{12}$, $I_{23}$, and $I_{31}$ shift the superconducting phases, controlling the position of the MZMs. A certain closed path in the space of the currents then implements an exchange of $\gamma_a$ and $\gamma_b$.
The configurations of the three junctions at three points along this path are shown in panels 1--3.
(b) The phases $\varphi_{1,2,3}(x)$ corresponding to the configuration shown in panel (a.1). 
The critical values $\varphi_{\pm}$ are indicated; the system is in the trivial phase when $\varphi_-<\varphi<\varphi_+$, and in the topological phase otherwise.
(c) The plane $\varphi_{1,0}+\varphi_{2,0}+\varphi_{3,0}=0$ to which the system is confined.
$\varphi_{i,0}$ are controlled by the externally injected currents, according to Eq.~(\ref{phaseconf}).
At the blue, green, and red lines, $\varphi_{1,0}$, $\varphi_{2,0}$, or $\varphi_{3,0}$, respectively cross one of the transition values ($\varphi_+$ or $\varphi_-$), either at the inner or outer edge of the corresponding junction.
The transition lines cut the plane into polygons, where each polygon hosts a fixed number of Majorana zero modes. The polygons are colored according to that number. The braiding path, indicated by red arrows, connects the three configurations 1--3 shown in panel (a). The left (right) white star marks the point where MZMs in junction 2 (1) are both at the outer edge, respectively.}
\label{fig:braiding}
\end{center}
\end{figure*}

The SNS junction we deal with has a different
potential $V(\varphi)$. This potential is affected by the parallel field $B_{||}$, due to the effect of $B_{||}$ on the Andreev spectrum. Generally, $V(\varphi)$ has local minima $\varphi_{1,2}$ [see Fig.~\ref{fig:system}(d)]. In the limit where the spin-orbit energy is much larger than the Zeeman energy these points are $\varphi_{1,2}\approx 0,\pi$~\cite{Pientka2017topological}. The parallel field $B_{||}$ determines which of the two is the global minimum. The soliton then starts and ends with $\varphi$ at the global minimum, but acquires a region where $\varphi$ is near the local minimum, and varies slowly [see Fig.~\ref{fig:system}(c)]. Remarkably, at the critical magnetic field $B_{||,c}$ where the potential $V(\varphi)$ has two degenerate minima $V(\varphi_1)=V(\varphi_2)$ the vortex splits to two half-vortices, each carrying a flux $hc/4e$. Since each half vortex spans a phase range of $\pi$, each will carry one MZM. Away from $B_{||,c}$
vortices are $2\pi$-vortices. As such they go through one pair of $\varphi_\pm$ values, and hence carry two MZMs, localized again at the points $x_\pm$. Close to the transition the separation between the two MZMs is of the order of $\lambda_J \log(B_{\parallel,c}/|B_\parallel - B_{\parallel,c}|)$, while far from the transition it approaches $\lambda_J$. In both cases, their spatial extent is $\xi_M=\max{\left (\xi_M^0,\sqrt{\xi_M^0\lambda_J}\right )}$. The coupling between the two MZMs is then a function of the ratio of $\xi_M$ and $\lambda_J$, and is not guaranteed to be small.

{\it Braiding scheme.---}Aiming towards a scheme for braiding, we now come back to the case of $\lambda_J\gg L$, where the coupling between MZMs may be better controlled.
Braiding of zero modes requires going beyond one dimension~\cite{Alicea2011non,Clarke2011majorana}. To that end, we consider a tri-junction geometry shown in Fig.~ \ref{fig:braiding}(a). The coordinates along the three junctions are $0<x_i<L$ ($i=1,2,3$). The three junctions meet at one point, $x_i=0$.

 The phase configuration of the tri-junction is determined by three equations of the form (\ref{current-configuraion}), augmented by the following boundary conditions:
\begin{align}
& \varphi_1(x_1=0)+\varphi_2(x_2=0)+\varphi_3(x_3=0)=2\pi n, \label{winding}\\
& \partial_x\varphi_1(x_1=0)=\partial_x\varphi_2(x_2=0)=\partial_x\varphi_3(x_3=0),
\label{centerfield}\\
& \frac{2\pi \kappa}{\Phi_0} \left[\partial_x\varphi_i(x_i=L)-\partial_x\varphi_i(x_i=0)\right]=\tilde{I}_{i} .
\end{align}
Here, $\tilde{I}_i$ is the current flowing through the $i$th junction. The phase of the superconductors can wind by $2\pi n$ around the point $x_i=0$, where $n$ is an integer. Since $\varphi$ denotes the phase-{\it difference} between the two superconductors that form a junction, the $2\pi n$ phase winding translates to the condition (\ref{winding}). The perpendicular magnetic field is continuous at the meeting point $x_i=0$, leading to the two equations (\ref{centerfield}). Finally, the imposition of the current through the junctions enforces the last equation. Note that the three currents through the junctions, $\tilde{I}_i$, are not independently controllable. Rather, by contacting the three superconductors in the junction we may control the three current differences $I_{i,i+1} = \tilde{I}_i-\tilde{I}_{i+1}$, out of which only two are independent [see Fig.~\ref{fig:braiding}(a.1)]. The currents $\tilde{I}_i$ may also include a circulating diamagnetic component which is not controlled by contacts.

In the presence of a perpendicular magnetic field, the phase varies linearly with the position along the junctions, according to Eq.~(\ref{no-screening})
[Fig.~\ref{fig:braiding}(b)]. We focus on the case $n=0$ in (\ref{winding}), in which the three phases $\varphi_i(x=0)$ which we denote by $\varphi_{i,0}$ are confined the plane $\sum_{i=1..3}\varphi_{i,0}=0$ [see Fig.~\ref{fig:braiding}(c)].

We shall label the phase configuration of the $i$th junction by a pair of binary digits $\chi_i = \{\chi^i_e,\chi^i_c\}$, according to whether the exterior end point, $x_i=L$, and the central point, $x_i=0$, are in the topological ($\chi^i_{e,c}=1$) or in the trivial ($\chi^i_{e,c}=0$) phase. If the magnetic field is weak enough, we are guaranteed that if $\chi^i_{e,c}$ are either both $0$ or both $1$, then there are no topological phase transitions in the $i$th junction.
Thus, the six binary digits $\chi_e^i,\chi_c^i$ determine the number of Majorana zero modes and their position. When $\chi_e^i=1$ the $i$th junction hosts a zero mode at $x_i=L$. When $\chi_e^i\ne \chi_c^i$ the $i$th junction hosts a Majorana mode somewhere between its two ends. If $\sum_i\chi_c^i$ is odd then there is a zero mode at the central point, $x_i=0$. Fig.~\ref{fig:braiding}(b) exemplifies two cases: one where both ends are trivial ($\chi_3 = \{0,0\}$), and another where the exterior end point is topological, while the central point is trivial ($\chi_{1,2}=\{1,0\}$).

Altogether, under these conditions the system may host zero, two, four or six Majorana modes (a larger number of Majorana modes requires a larger perpendicular field, such that several transitions may take place along one junction). Fig.~\ref{fig:braiding}(c) is colored according to the number of Majorana modes hosted by the system as a function of the phases at the center points, $\varphi_{i,0}$. At the value of the perpendicular field chosen in the figure there is no region where all junctions are trivial. Such a region can occur for weaker fields.

Motion within the plane in Fig.~\ref{fig:braiding}(c) is driven by currents. The number of Majorana modes may vary when such motion changes $\chi^i_{c}$ or $\chi^i_{e}$. In particular, a transition from $\chi_i =\{0,0\}$ to $\{1,0\}$ indicates the creation of two Majorana modes at the $i$th junction, initialized in the vacuum state.

To perform braiding, we need to have at least four Majorana modes. A smaller number does not allow for non-abelian unitary transformations (since the overall fermion parity is fixed).
A braiding manipulation corresponds to a closed trajectory in the plane presented in Fig.~\ref{fig:braiding}(c). The trajectory should be non-contractable to a point, that is, it should encircle a region where the number of Majorana modes is different from four.

An example to such a trajectory is shown in panel (c) of Fig.~\ref{fig:braiding}. It is elaborated on in panels (a1--a3) of the Figure, and to greater details (including animation) in the Supplementary Matreial. It begins with $\chi_1=\chi_2=\{1,0\}$ and $\chi_3=\{0,0\}$.  The system then hosts four Majorana modes, which we denote by $\gamma_a,\gamma_b,\gamma_e,\gamma_f$. It is useful to regard the central point $x_i=0$ as hosting two additional Majorana modes, $\gamma_c,\gamma_d$, that are strongly coupled to each other. Moving the system to point (2) in panel (c), across a transition line in which $\chi_2,\chi_3$ change to $\{1,1\},\{0,1\}$, respectively, leads to the situation depicted in panel (a.2), where the modes $\gamma_b,\gamma_c$ are coupled, while $\gamma_d$ is a zero mode. Next, moving the system to point 3 in panel (c) we change $\chi_2,\chi_3$ to $\{0,1\}, \{1,1\}$, respectively (panel (a.3)); then, $\gamma_a,\gamma_c$ are strongly coupled, and $\gamma_b$ is a zero mode. The braiding is then completed by going back to point (1), panel (a.1). {This process effectively interchanges $\gamma_a$ and $\gamma_b$, and is described by the action of the unitary operator $e^{\frac{\pi}{4} \gamma_a \gamma_b}$ on the ground state subspace.}

{The same setup allows also to initialize the system in a certain state and measure the outcome of the braiding. To initialize a pair of zero modes in a given state, they can be nucleated from the vacuum; for example, it is possible to tune the phases such that junction number 1 is entirely in the trivial state, ($\chi_1 = \{0,0\}$), and then the Majorana modes $\gamma_b$, $\gamma_f$ are pushed to the outer end of the junction and are strongly coupled to each other. This situation is realized at the white star labeled as $R$ in Fig.~\ref{fig:braiding}(c). If the system is then coupled to a metallic lead, $\gamma_b$, $\gamma_f$ are initialized in their joint ground state (e.g., $i\gamma_b \gamma_f = 1$). Tuning the phases back to the point labelled as 1 in Fig.~\ref{fig:braiding}(c) ($\chi_1=\{1,0\}$) decouples $\gamma_b$ and $\gamma_f$, bringing them back to zero energy. Similarly, $\gamma_a$, $\gamma_e$ can be initialized by tuning the phases to the white star labeled as $L$ in Fig.~\ref{fig:braiding}(c).}

The same process that allows initializing $\gamma_b$, $\gamma_f$ allows also to measure their joint parity after the braiding process (exchanging $\gamma_a$, $\gamma_b$) is complete. This can be done by bringing $\gamma_b$, $\gamma_f$ to the end of junction 1, which removes the degeneracy by coupling them to one another, and then coupling the junction's end to a quantum dot. The current from the dot to the junction may then measure the fermion occupation~\cite{Plugge2017majorana,Karzig2017scalable}.

In summary, in this paper we examined the effect of a perpendicular magnetic field and a driving current on one-dimensional topological superconductors formed at the normal part of an SNS Josephson junction in the presence of spin-orbit coupling and a parallel magnetic field. In particular, we demonstrated the fractionalization of Josephson vortices
and the possibility of current-controlled braiding of Majorana zero modes in this setup.

{\it Acknowledgements.---}A.~S. and E.~B. acknowledge support from CRC 183 of the Deutsche Forschungsgemeinschaft. E.~B. is grateful to the Aspen Center for Physics, were part of this work was done. A.~S. is supported by the European Research Council (Project MUNATOP), by the Israel Science Foundation and by Microsoft's Station Q.

\bibliography{References}

\clearpage
\widetext
\input{SO-JJ-half-vortex-supp_arxiv-submission.tex}

\end{document}

%% file: SO-JJ-half-vortex-supp_arxiv-submission.tex
%
%
%
%
%


\makeatother


%

\maketitle

\section*{Supplementary Material}

\setcounter{figure}{0}
\renewcommand{\thefigure}{S\arabic{figure}}

\setcounter{equation}{0}
\renewcommand{\theequation}{S\arabic{equation}}

\subsection{Further details on the braiding process}

\begin{figure}[b]
\begin{center}
\includegraphics[width=0.6\textwidth]{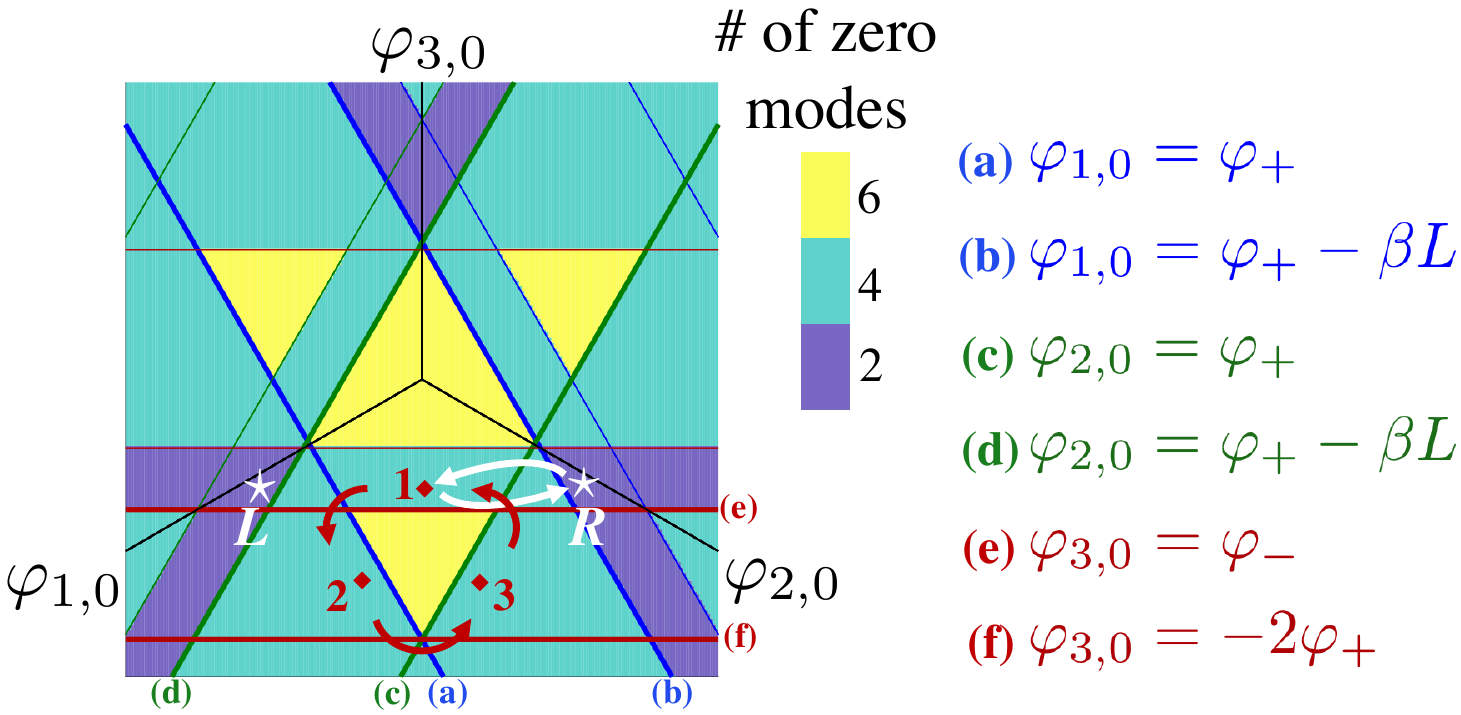}
\caption{The plane $\varphi_{1,0}+\varphi_{2,0}+\varphi_{3,0}=0$ in which the braiding takes place. Special lines of constant $\varphi_{1,0},\varphi_{2,0},\varphi_{3,0}$ are indicated by blue, green, and red lines; the corresponding values of the phases at the thick lines marked by (a)-(f) are shown on the right. In order to perform the initialization, braiding, and measurement process, the three superconducting phases $\varphi_{1,0},\varphi_{2,0},\varphi_{3,0}$ need to be controlled at least within the range bounded by the pairs of lines (a,b), (c,d), and (e,f), respectively.}
\label{fig:braiding_sup}
\end{center}
\end{figure}

Here, we provide additional details on the proposed tri-junction setup and the protocol for initializing and braiding MZMs. As discussed in the main text, the initialization and braiding process can be described as a trajectory in the $\varphi_{1,0}+\varphi_{2,0}+\varphi_{3,0}=0$ plane, shown in Fig.~\ref{fig:braiding_sup}, where we also indicate special lines of constant $\varphi_{i,0}$ which play an important role in the process. An animation showing the braiding process, both in real space and in the space of $\{\varphi_{i,0}\}$, can be found here: \url{https://www.dropbox.com/s/ps4hhfwvf1k483y/braiding_movie.avi?dl=0}~\cite{youtube}. We start the process from a point labeled 1 in the plane. We then initialize the Majorana zero modes $\gamma_b$ and $\gamma_f$ to a state of well defined parity, $i\gamma_b \gamma_f = 1$. This is done by moving to the point labeled as $R$ in Fig.~\ref{fig:braiding_sup}; at that point, the entire junction 1 is in the trivial phase, and $\gamma_b$ and $\gamma_f$ are strongly coupled. We then move to the point labeled by $L$, where $\gamma_a$, $\gamma_e$ are initialized to a state of well-defined parity $i\gamma_a \gamma_e = 1$. The braiding process then consists of moving around the trajectory $1\rightarrow 2 \rightarrow 3 \rightarrow 1$, which interchanges $\gamma_a$ and $\gamma_b$. The joint parity $i \gamma_a \gamma_e$ can then be measured; by the non-Abelian braiding rules for MZMs, the system is in an equal superposition of the states $i\gamma_a \gamma_e = \pm1$, and the result of the measurement can either be $+1$ or $-1$ with a $50\%$ probability~\cite{Ivanov2001non}.

This protocol requires the ability to control the phases $\varphi_{i,0}$ within a minimal range. The parameters $B_\perp$, $B_\parallel$, and $L$ should be chosen such that the trajectory in Fig.~\ref{fig:braiding_sup} can be covered. For example, during the process, $\varphi_{3,0}$ needs to vary at least within the regime $[-2\varphi_+, \varphi_-]$, where $\varphi_{\pm}$ are functions of $B_\parallel$~\cite{Pientka2017topological}. The range of accessible $\varphi_{i,0}$ is determined by the energy-phase relation of the junction.  To illustrate how this range is calculated, it is useful to consider the case where the energy-phase relation of the junction is sinusoidal, such that the potential in Eq.~(1) of the main text is given by $V(\varphi) = \epsilon_J (1 - \cos\varphi)$; in that case, by Eq.~(3) of the main text, the current through the junction is given by
\begin{equation}
\tilde{I}_i = \frac{2\pi}{\phi_0} \int_0^L dx V'\big(\varphi(x)\big) = \frac{2\pi}{\phi_0} \epsilon_J \frac{\cos(\varphi_{i,0}) - \cos(\varphi_{i,0}+\beta L) }{\beta} =
I_c \sin(\varphi_{i,0} + \frac{\beta L}{2}) ,
\end{equation}
where $I_c = \frac{4\pi}{\phi_0} \epsilon_J \frac{\sin(\frac{\beta L}{2})}{\beta}$ is the critical current of the junction. Thus, tuning the current between $-I_c$ and $I_c$ allows to set the phase $\varphi_{i,0}$ to any value in the range $\left[-\frac{\beta L}{2} - \frac{\pi}{2}, -\frac{\beta L}{2} + \frac{\pi}{2}\right]$. The range of accessible $\varphi_{i,0}$ can be calculated for a more complicated energy-phase relation in a similar fashion.

Another requirement is that the energy gap in the junctions is sufficiently large, such that the MZM localization length is much shorter than $L$. For this purpose, it is important that the three junctions are parallel to the direction of the in-plane magnetic field over most of their length~\cite{Pientka2017topological}, as in Fig.~1(a). In addition, distance between the junctions in junctions 1 and 2 of the device is required to be larger than the bulk superconducting coherence length.

\begin{figure}[t]
\begin{center}
\includegraphics[width=0.4\textwidth]{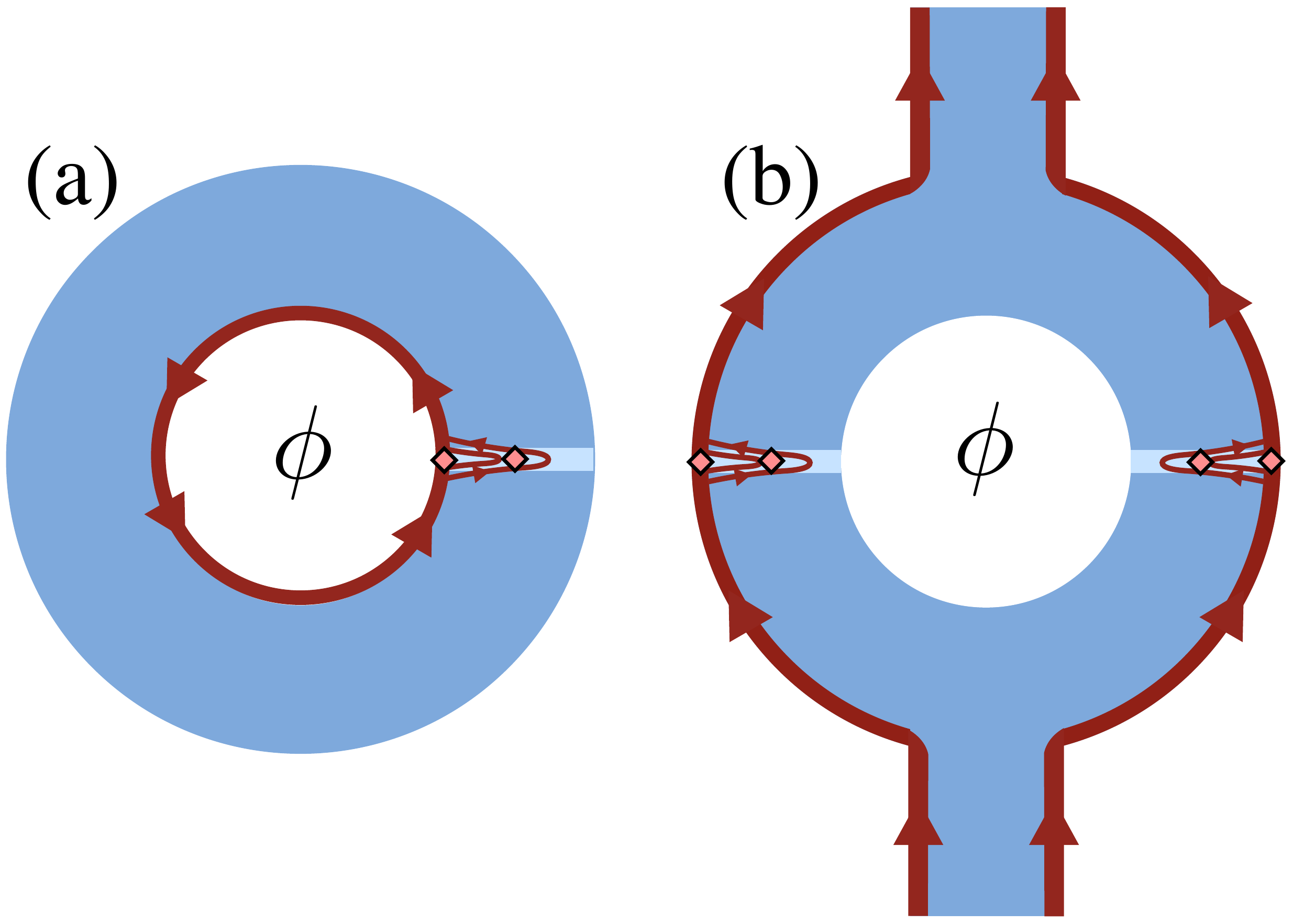}
\caption{The two geometries discussed in the text: (a) annular superconductor geometry, (b) SQUID geometry. In both geometries, a flux $\phi$ is threaded through the hole at the center. The bright region denotes the junction. The current path is shown in arrows, and the locations of the MZMs are indicated by diamonds. The distance between the MZMs is of the order of $\lambda_J$ in both geometries. The SQUID drawn has two identical junctions, but that does not have to be the case.}
\label{fig:geometries}
\end{center}
\end{figure}

\subsection{The position of MZMs in the presence of screening currents}

So far, we mostly focused on the limit where the junction is much shorter than $\lambda_J$, such that screening currents may be neglected. Here, we comment on the position of the MZMs in cases where the junction is long. For simplicity, we initially assume that no perpendicular magnetic field is externally applied (although such a field may be created by the screening currents). We distinguish between two geometries - the flux loop and the current-driven SQUID. The flux loop is made of an annular superconductor that encloses a Josephson junction, with a flux $|\phi|<\phi_0/2$ threading at the center of the annulus (see Fig.~\ref{fig:geometries}). Were the loop made of a uniform superconductor, the flux would have been screened by a screening current flowing at the interior side of the annulus, within a London distance $\lambda_L$ from the edge. within the Josephson junction, that distance is replaced by  $\lambda_J$, which is typically much larger than $\lambda_L$.  The phase configuration within the junction is determined by Eq.~(1) of the main text, with $\varphi(x=0)=2\pi\phi/\phi_0$ and $\partial_x\varphi(x=L)=0$. For $L\gg\lambda_J$, the phase evolves from $2\pi\phi/\phi_0$ at the interior edge to $\varphi_{min}$, the minimum point of $V(\varphi)$. For small values of $B_{||}$, before the first order phase transition, $\varphi_{min}\approx 0$. When $\varphi_-<\varphi(x=0)<\varphi_+$, part of the junction is in a topological state, and there is one MZM centered near $x=0$ and another one centered at the point where the phase takes the value  $\varphi(x)=\varphi_-$, on its way to the asymptotic value $\varphi_{min}$.

The situation is different in a SQUID geometry, where current is driven through two junctions of an interference loop, and each junction encloses a Josephson junction. As long as the arms of the SQUID are wider than $\lambda_L$, the current flows within $\lambda_L$ of the exterior side of the two arms (see Fig.~\ref{fig:geometries}). When the current crosses the junctions, it spreads into a distance of $\lambda_J$ away from the exterior edge of the junction. Thus, in this geometry the phase varies close to the exterior edge, and takes the value $\varphi_{min}$ at distances larger than $\lambda_J$ from that edge. The value of the phase at $x=L$ is determined by the current driven through the SQUID. When that value satisfies $\varphi_- < \varphi(x=L) < \varphi_+$ part of the junction is in the topological phase, one MZM is at $x=L$ and another is where $\varphi(x)=\varphi_-$.

In either geometry, when the flux {\it through} the junction is larger than one flux quantum, the junction carries Josephson vortices. The properties of these vortices and the positions taken by the MZMs they carry are described in the main text.

%

%% file: SO-JJ-half-vortex_arxiv-submission.bbl
\begin{thebibliography}{32}%
\makeatletter
\providecommand \@ifxundefined [1]{%
 \@ifx{#1\undefined}
}%
\providecommand \@ifnum [1]{%
 \ifnum #1\expandafter \@firstoftwo
 \else \expandafter \@secondoftwo
 \fi
}%
\providecommand \@ifx [1]{%
 \ifx #1\expandafter \@firstoftwo
 \else \expandafter \@secondoftwo
 \fi
}%
\providecommand \natexlab [1]{#1}%
\providecommand \enquote  [1]{``#1''}%
\providecommand \bibnamefont  [1]{#1}%
\providecommand \bibfnamefont [1]{#1}%
\providecommand \citenamefont [1]{#1}%
\providecommand \href@noop [0]{\@secondoftwo}%
\providecommand \href [0]{\begingroup \@sanitize@url \@href}%
\providecommand \@href[1]{\@@startlink{#1}\@@href}%
\providecommand \@@href[1]{\endgroup#1\@@endlink}%
\providecommand \@sanitize@url [0]{\catcode `\\12\catcode `\$12\catcode
  `\&12\catcode `\#12\catcode `\^12\catcode `\_12\catcode `\%12\relax}%
\providecommand \@@startlink[1]{}%
\providecommand \@@endlink[0]{}%
\providecommand \url  [0]{\begingroup\@sanitize@url \@url }%
\providecommand \@url [1]{\endgroup\@href {#1}{\urlprefix }}%
\providecommand \urlprefix  [0]{URL }%
\providecommand \Eprint [0]{\href }%
\providecommand \doibase [0]{http://dx.doi.org/}%
\providecommand \selectlanguage [0]{\@gobble}%
\providecommand \bibinfo  [0]{\@secondoftwo}%
\providecommand \bibfield  [0]{\@secondoftwo}%
\providecommand \translation [1]{[#1]}%
\providecommand \BibitemOpen [0]{}%
\providecommand \bibitemStop [0]{}%
\providecommand \bibitemNoStop [0]{.\EOS\space}%
\providecommand \EOS [0]{\spacefactor3000\relax}%
\providecommand \BibitemShut  [1]{\csname bibitem#1\endcsname}%
\let\auto@bib@innerbib\@empty
\bibitem [{\citenamefont {Lutchyn}\ \emph {et~al.}(2018)\citenamefont
  {Lutchyn}, \citenamefont {Bakkers}, \citenamefont {Kouwenhoven},
  \citenamefont {Krogstrup}, \citenamefont {Marcus},\ and\ \citenamefont
  {Oreg}}]{Lutchyn2018majorana}%
  \BibitemOpen
  \bibfield  {author} {\bibinfo {author} {\bibfnamefont {RM}~\bibnamefont
  {Lutchyn}}, \bibinfo {author} {\bibfnamefont {EPAM}\ \bibnamefont {Bakkers}},
  \bibinfo {author} {\bibfnamefont {LP}~\bibnamefont {Kouwenhoven}}, \bibinfo
  {author} {\bibfnamefont {P}~\bibnamefont {Krogstrup}}, \bibinfo {author}
  {\bibfnamefont {CM}~\bibnamefont {Marcus}}, \ and\ \bibinfo {author}
  {\bibfnamefont {Y}~\bibnamefont {Oreg}},\ }\bibfield  {title} {\enquote
  {\bibinfo {title} {Majorana zero modes in superconductor--semiconductor
  heterostructures},}\ }\href
  {https://www.nature.com/articles/s41578-018-0003-1} {\bibfield  {journal}
  {\bibinfo  {journal} {Nat. Rev. Mat.}\ }\textbf {\bibinfo {volume} {3}},\
  \bibinfo {pages} {52--68} (\bibinfo {year} {2018})}\BibitemShut {NoStop}%
\bibitem [{\citenamefont {Aguado}()}]{Aguado2017majorana}%
  \BibitemOpen
  \bibfield  {author} {\bibinfo {author} {\bibfnamefont {Ram{\'o}n}\
  \bibnamefont {Aguado}},\ }\bibfield  {title} {\enquote {\bibinfo {title}
  {Majorana quasiparticles in condensed matter},}\ }\href@noop {} {\bibinfo
  {journal} {Riv. Nuovo Cimento}\ }\BibitemShut {NoStop}%
\bibitem [{\citenamefont {Alicea}(2012)}]{Alicea2012}%
  \BibitemOpen
\bibfield  {journal} {  }\bibfield  {author} {\bibinfo {author} {\bibfnamefont
  {Jason}\ \bibnamefont {Alicea}},\ }\bibfield  {title} {\enquote {\bibinfo
  {title} {{New directions in the pursuit of Majorana fermions in solid state
  systems.}}}\ }\href {\doibase 10.1088/0034-4885/75/7/076501} {\bibfield
  {journal} {\bibinfo  {journal} {Rep. Prog. Phys.}\ }\textbf {\bibinfo
  {volume} {75}},\ \bibinfo {pages} {076501} (\bibinfo {year}
  {2012})}\BibitemShut {NoStop}%
\bibitem [{\citenamefont {Beenakker}(2013)}]{Beenakker2013}%
  \BibitemOpen
  \bibfield  {author} {\bibinfo {author} {\bibfnamefont {C.~W.~J.}\
  \bibnamefont {Beenakker}},\ }\bibfield  {title} {\enquote {\bibinfo {title}
  {{Search for Majorana Fermions in Superconductors}},}\ }\href {\doibase
  10.1146/annurev-conmatphys-030212-184337} {\bibfield  {journal} {\bibinfo
  {journal} {Ann. Rev. Condens. Matt. Phys.}\ }\textbf {\bibinfo {volume}
  {4}},\ \bibinfo {pages} {113--136} (\bibinfo {year} {2013})}\BibitemShut
  {NoStop}%
\bibitem [{\citenamefont {Kopnin}\ and\ \citenamefont
  {Salomaa}(1991)}]{Kopnin1991mutual}%
  \BibitemOpen
  \bibfield  {author} {\bibinfo {author} {\bibfnamefont {N.~B.}\ \bibnamefont
  {Kopnin}}\ and\ \bibinfo {author} {\bibfnamefont {M.~M.}\ \bibnamefont
  {Salomaa}},\ }\bibfield  {title} {\enquote {\bibinfo {title} {Mutual friction
  in superfluid $^{3}\mathrm{He}$: Effects of bound states in the vortex
  core},}\ }\href {\doibase 10.1103/PhysRevB.44.9667} {\bibfield  {journal}
  {\bibinfo  {journal} {Phys. Rev. B}\ }\textbf {\bibinfo {volume} {44}},\
  \bibinfo {pages} {9667--9677} (\bibinfo {year} {1991})}\BibitemShut {NoStop}%
\bibitem [{\citenamefont {Read}\ and\ \citenamefont
  {Green}(2000)}]{Read2000paired}%
  \BibitemOpen
  \bibfield  {author} {\bibinfo {author} {\bibfnamefont {N.}~\bibnamefont
  {Read}}\ and\ \bibinfo {author} {\bibfnamefont {D.}~\bibnamefont {Green}},\
  }\bibfield  {title} {\enquote {\bibinfo {title} {Paired states of fermions in
  two dimensions with breaking of parity and time-reversal symmetries and the
  fractional quantum hall effect},}\ }\href
  {http://link.aps.org/doi/10.1103/PhysRevB.61.10267} {\bibfield  {journal}
  {\bibinfo  {journal} {Phys. Rev. B}\ }\textbf {\bibinfo {volume} {61}},\
  \bibinfo {pages} {10267} (\bibinfo {year} {2000})}\BibitemShut {NoStop}%
\bibitem [{\citenamefont {Kitaev}(2001)}]{Kitaev2001unpaired}%
  \BibitemOpen
  \bibfield  {author} {\bibinfo {author} {\bibfnamefont {A.Y.}\ \bibnamefont
  {Kitaev}},\ }\bibfield  {title} {\enquote {\bibinfo {title} {Unpaired
  majorana fermions in quantum wires},}\ }\href
  {http://iopscience.iop.org/article/10.1070/1063-7869/44/10S/S29/meta}
  {\bibfield  {journal} {\bibinfo  {journal} {Phys. Usp.}\ }\textbf {\bibinfo
  {volume} {44}},\ \bibinfo {pages} {131} (\bibinfo {year} {2001})}\BibitemShut
  {NoStop}%
\bibitem [{\citenamefont {Lutchyn}\ \emph {et~al.}(2010)\citenamefont
  {Lutchyn}, \citenamefont {Sau},\ and\ \citenamefont
  {Das~Sarma}}]{Lutchyn2010majorana}%
  \BibitemOpen
  \bibfield  {author} {\bibinfo {author} {\bibfnamefont {Roman~M.}\
  \bibnamefont {Lutchyn}}, \bibinfo {author} {\bibfnamefont {Jay~D.}\
  \bibnamefont {Sau}}, \ and\ \bibinfo {author} {\bibfnamefont
  {S.}~\bibnamefont {Das~Sarma}},\ }\bibfield  {title} {\enquote {\bibinfo
  {title} {Majorana fermions and a topological phase transition in
  semiconductor-superconductor heterostructures},}\ }\href {\doibase
  10.1103/PhysRevLett.105.077001} {\bibfield  {journal} {\bibinfo  {journal}
  {Phys. Rev. Lett.}\ }\textbf {\bibinfo {volume} {105}},\ \bibinfo {pages}
  {077001} (\bibinfo {year} {2010})}\BibitemShut {NoStop}%
\bibitem [{\citenamefont {Sau}\ \emph {et~al.}(2010)\citenamefont {Sau},
  \citenamefont {Lutchyn}, \citenamefont {Tewari},\ and\ \citenamefont
  {Das~Sarma}}]{Sau2010generic}%
  \BibitemOpen
  \bibfield  {author} {\bibinfo {author} {\bibfnamefont {Jay~D.}\ \bibnamefont
  {Sau}}, \bibinfo {author} {\bibfnamefont {Roman~M.}\ \bibnamefont {Lutchyn}},
  \bibinfo {author} {\bibfnamefont {Sumanta}\ \bibnamefont {Tewari}}, \ and\
  \bibinfo {author} {\bibfnamefont {S.}~\bibnamefont {Das~Sarma}},\ }\bibfield
  {title} {\enquote {\bibinfo {title} {Generic new platform for topological
  quantum computation using semiconductor heterostructures},}\ }\href {\doibase
  10.1103/PhysRevLett.104.040502} {\bibfield  {journal} {\bibinfo  {journal}
  {Phys. Rev. Lett.}\ }\textbf {\bibinfo {volume} {104}},\ \bibinfo {pages}
  {040502} (\bibinfo {year} {2010})}\BibitemShut {NoStop}%
\bibitem [{\citenamefont {Oreg}\ \emph {et~al.}(2010)\citenamefont {Oreg},
  \citenamefont {Refael},\ and\ \citenamefont {von Oppen}}]{Oreg2010helical}%
  \BibitemOpen
  \bibfield  {author} {\bibinfo {author} {\bibfnamefont {Yuval}\ \bibnamefont
  {Oreg}}, \bibinfo {author} {\bibfnamefont {Gil}\ \bibnamefont {Refael}}, \
  and\ \bibinfo {author} {\bibfnamefont {Felix}\ \bibnamefont {von Oppen}},\
  }\bibfield  {title} {\enquote {\bibinfo {title} {Helical liquids and majorana
  bound states in quantum wires},}\ }\href {\doibase
  10.1103/PhysRevLett.105.177002} {\bibfield  {journal} {\bibinfo  {journal}
  {Phys. Rev. Lett.}\ }\textbf {\bibinfo {volume} {105}},\ \bibinfo {pages}
  {177002} (\bibinfo {year} {2010})}\BibitemShut {NoStop}%
\bibitem [{\citenamefont {Nadj-Perge}\ \emph {et~al.}(2013)\citenamefont
  {Nadj-Perge}, \citenamefont {Drozdov}, \citenamefont {Bernevig},\ and\
  \citenamefont {Yazdani}}]{Nadj2013proposal}%
  \BibitemOpen
  \bibfield  {author} {\bibinfo {author} {\bibfnamefont {S}~\bibnamefont
  {Nadj-Perge}}, \bibinfo {author} {\bibfnamefont {IK}~\bibnamefont {Drozdov}},
  \bibinfo {author} {\bibfnamefont {BA}~\bibnamefont {Bernevig}}, \ and\
  \bibinfo {author} {\bibfnamefont {Ali}\ \bibnamefont {Yazdani}},\ }\bibfield
  {title} {\enquote {\bibinfo {title} {Proposal for realizing majorana fermions
  in chains of magnetic atoms on a superconductor},}\ }\href@noop {} {\bibfield
   {journal} {\bibinfo  {journal} {Physical Review B}\ }\textbf {\bibinfo
  {volume} {88}},\ \bibinfo {pages} {020407} (\bibinfo {year}
  {2013})}\BibitemShut {NoStop}%
\bibitem [{\citenamefont {Klinovaja}\ \emph {et~al.}(2013)\citenamefont
  {Klinovaja}, \citenamefont {Stano}, \citenamefont {Yazdani},\ and\
  \citenamefont {Loss}}]{Klinovaja2013topological}%
  \BibitemOpen
  \bibfield  {author} {\bibinfo {author} {\bibfnamefont {Jelena}\ \bibnamefont
  {Klinovaja}}, \bibinfo {author} {\bibfnamefont {Peter}\ \bibnamefont
  {Stano}}, \bibinfo {author} {\bibfnamefont {Ali}\ \bibnamefont {Yazdani}}, \
  and\ \bibinfo {author} {\bibfnamefont {Daniel}\ \bibnamefont {Loss}},\
  }\bibfield  {title} {\enquote {\bibinfo {title} {Topological
  superconductivity and majorana fermions in rkky systems},}\ }\href {\doibase
  10.1103/PhysRevLett.111.186805} {\bibfield  {journal} {\bibinfo  {journal}
  {Phys. Rev. Lett.}\ }\textbf {\bibinfo {volume} {111}},\ \bibinfo {pages}
  {186805} (\bibinfo {year} {2013})}\BibitemShut {NoStop}%
\bibitem [{\citenamefont {Mourik}\ \emph {et~al.}(2012)\citenamefont {Mourik},
  \citenamefont {Zuo}, \citenamefont {Frolov}, \citenamefont {Plissard},
  \citenamefont {Bakkers},\ and\ \citenamefont
  {Kouwenhoven}}]{Mourik2012signatures}%
  \BibitemOpen
  \bibfield  {author} {\bibinfo {author} {\bibfnamefont {V.}~\bibnamefont
  {Mourik}}, \bibinfo {author} {\bibfnamefont {K.}~\bibnamefont {Zuo}},
  \bibinfo {author} {\bibfnamefont {SM}~\bibnamefont {Frolov}}, \bibinfo
  {author} {\bibfnamefont {SR}~\bibnamefont {Plissard}}, \bibinfo {author}
  {\bibfnamefont {E.}~\bibnamefont {Bakkers}}, \ and\ \bibinfo {author}
  {\bibfnamefont {LP}~\bibnamefont {Kouwenhoven}},\ }\bibfield  {title}
  {\enquote {\bibinfo {title} {Signatures of majorana fermions in hybrid
  superconductor-semiconductor nanowire devices},}\ }\href
  {http://www.sciencemag.org/content/336/6084/1003} {\bibfield  {journal}
  {\bibinfo  {journal} {Science}\ }\textbf {\bibinfo {volume} {336}},\ \bibinfo
  {pages} {1003--1007} (\bibinfo {year} {2012})}\BibitemShut {NoStop}%
\bibitem [{\citenamefont {Das}\ \emph {et~al.}(2012)\citenamefont {Das},
  \citenamefont {Ronen}, \citenamefont {Most}, \citenamefont {Oreg},
  \citenamefont {Heiblum},\ and\ \citenamefont {Shtrikman}}]{Das2012zero}%
  \BibitemOpen
  \bibfield  {author} {\bibinfo {author} {\bibfnamefont {Anindya}\ \bibnamefont
  {Das}}, \bibinfo {author} {\bibfnamefont {Yuval}\ \bibnamefont {Ronen}},
  \bibinfo {author} {\bibfnamefont {Yonatan}\ \bibnamefont {Most}}, \bibinfo
  {author} {\bibfnamefont {Yuval}\ \bibnamefont {Oreg}}, \bibinfo {author}
  {\bibfnamefont {Moty}\ \bibnamefont {Heiblum}}, \ and\ \bibinfo {author}
  {\bibfnamefont {Hadas}\ \bibnamefont {Shtrikman}},\ }\bibfield  {title}
  {\enquote {\bibinfo {title} {{Zero-bias peaks and splitting in an AlגAlInAs
  nanowire topological superconductor as a signature of Majorana fermions}},}\
  }\href {\doibase 10.1038/nphys2479} {\bibfield  {journal} {\bibinfo
  {journal} {Nat. Phys.}\ }\textbf {\bibinfo {volume} {8}},\ \bibinfo {pages}
  {887} (\bibinfo {year} {2012})}\BibitemShut {NoStop}%
\bibitem [{\citenamefont {Albrecht}\ \emph {et~al.}(2016)\citenamefont
  {Albrecht}, \citenamefont {Higginbotham}, \citenamefont {Madsen},
  \citenamefont {Kuemmeth}, \citenamefont {Jespersen}, \citenamefont
  {Nyg{\aa}rd}, \citenamefont {Krogstrup},\ and\ \citenamefont
  {Marcus}}]{Albrecht2016exponential}%
  \BibitemOpen
  \bibfield  {author} {\bibinfo {author} {\bibfnamefont {SM}~\bibnamefont
  {Albrecht}}, \bibinfo {author} {\bibfnamefont {AP}~\bibnamefont
  {Higginbotham}}, \bibinfo {author} {\bibfnamefont {M}~\bibnamefont {Madsen}},
  \bibinfo {author} {\bibfnamefont {F}~\bibnamefont {Kuemmeth}}, \bibinfo
  {author} {\bibfnamefont {TS}~\bibnamefont {Jespersen}}, \bibinfo {author}
  {\bibfnamefont {Jesper}\ \bibnamefont {Nyg{\aa}rd}}, \bibinfo {author}
  {\bibfnamefont {P}~\bibnamefont {Krogstrup}}, \ and\ \bibinfo {author}
  {\bibfnamefont {CM}~\bibnamefont {Marcus}},\ }\bibfield  {title} {\enquote
  {\bibinfo {title} {Exponential protection of zero modes in majorana
  islands},}\ }\href
  {http://www.nature.com/nature/journal/v531/n7593/abs/nature17162.html}
  {\bibfield  {journal} {\bibinfo  {journal} {Nature}\ }\textbf {\bibinfo
  {volume} {531}},\ \bibinfo {pages} {206--209} (\bibinfo {year}
  {2016})}\BibitemShut {NoStop}%
\bibitem [{\citenamefont {Nadj-Perge}\ \emph {et~al.}(2014)\citenamefont
  {Nadj-Perge}, \citenamefont {Drozdov}, \citenamefont {Li}, \citenamefont
  {Chen}, \citenamefont {Jeon}, \citenamefont {Seo}, \citenamefont {MacDonald},
  \citenamefont {Bernevig},\ and\ \citenamefont
  {Yazdani}}]{Nadj-Perge2014observation}%
  \BibitemOpen
  \bibfield  {author} {\bibinfo {author} {\bibfnamefont {Stevan}\ \bibnamefont
  {Nadj-Perge}}, \bibinfo {author} {\bibfnamefont {Ilya~K.}\ \bibnamefont
  {Drozdov}}, \bibinfo {author} {\bibfnamefont {Jian}\ \bibnamefont {Li}},
  \bibinfo {author} {\bibfnamefont {Hua}\ \bibnamefont {Chen}}, \bibinfo
  {author} {\bibfnamefont {Sangjun}\ \bibnamefont {Jeon}}, \bibinfo {author}
  {\bibfnamefont {Jungpil}\ \bibnamefont {Seo}}, \bibinfo {author}
  {\bibfnamefont {Allan~H.}\ \bibnamefont {MacDonald}}, \bibinfo {author}
  {\bibfnamefont {B.~Andrei}\ \bibnamefont {Bernevig}}, \ and\ \bibinfo
  {author} {\bibfnamefont {Ali}\ \bibnamefont {Yazdani}},\ }\bibfield  {title}
  {\enquote {\bibinfo {title} {Observation of majorana fermions in
  ferromagnetic atomic chains on a superconductor},}\ }\href
  {http://www.sciencemag.org/content/early/2014/10/01/science.1259327.abstract}
  {\bibfield  {journal} {\bibinfo  {journal} {Science}\ }\textbf {\bibinfo
  {volume} {346}},\ \bibinfo {pages} {602} (\bibinfo {year}
  {2014})}\BibitemShut {NoStop}%
\bibitem [{\citenamefont {Shabani}\ \emph {et~al.}(2016)\citenamefont
  {Shabani}, \citenamefont {Kjaergaard}, \citenamefont {Suominen},
  \citenamefont {Kim}, \citenamefont {Nichele}, \citenamefont {Pakrouski},
  \citenamefont {Stankevic}, \citenamefont {Lutchyn}, \citenamefont
  {Krogstrup}, \citenamefont {Feidenhans} \emph {et~al.}}]{Shabani2016two}%
  \BibitemOpen
  \bibfield  {author} {\bibinfo {author} {\bibfnamefont {J}~\bibnamefont
  {Shabani}}, \bibinfo {author} {\bibfnamefont {M}~\bibnamefont {Kjaergaard}},
  \bibinfo {author} {\bibfnamefont {HJ}~\bibnamefont {Suominen}}, \bibinfo
  {author} {\bibfnamefont {Younghyun}\ \bibnamefont {Kim}}, \bibinfo {author}
  {\bibfnamefont {F}~\bibnamefont {Nichele}}, \bibinfo {author} {\bibfnamefont
  {K}~\bibnamefont {Pakrouski}}, \bibinfo {author} {\bibfnamefont
  {T}~\bibnamefont {Stankevic}}, \bibinfo {author} {\bibfnamefont {Roman~M}\
  \bibnamefont {Lutchyn}}, \bibinfo {author} {\bibfnamefont {P}~\bibnamefont
  {Krogstrup}}, \bibinfo {author} {\bibfnamefont {R}~\bibnamefont
  {Feidenhans}},  \emph {et~al.},\ }\bibfield  {title} {\enquote {\bibinfo
  {title} {Two-dimensional epitaxial superconductor-semiconductor
  heterostructures: A platform for topological superconducting networks},}\
  }\href@noop {} {\bibfield  {journal} {\bibinfo  {journal} {Physical Review
  B}\ }\textbf {\bibinfo {volume} {93}},\ \bibinfo {pages} {155402} (\bibinfo
  {year} {2016})}\BibitemShut {NoStop}%
\bibitem [{\citenamefont {Suominen}\ \emph {et~al.}(2017)\citenamefont
  {Suominen}, \citenamefont {Kjaergaard}, \citenamefont {Hamilton},
  \citenamefont {Shabani}, \citenamefont {Palmstr\o{}m}, \citenamefont
  {Marcus},\ and\ \citenamefont {Nichele}}]{Suominen2017zero}%
  \BibitemOpen
  \bibfield  {author} {\bibinfo {author} {\bibfnamefont {H.~J.}\ \bibnamefont
  {Suominen}}, \bibinfo {author} {\bibfnamefont {M.}~\bibnamefont
  {Kjaergaard}}, \bibinfo {author} {\bibfnamefont {A.~R.}\ \bibnamefont
  {Hamilton}}, \bibinfo {author} {\bibfnamefont {J.}~\bibnamefont {Shabani}},
  \bibinfo {author} {\bibfnamefont {C.~J.}\ \bibnamefont {Palmstr\o{}m}},
  \bibinfo {author} {\bibfnamefont {C.~M.}\ \bibnamefont {Marcus}}, \ and\
  \bibinfo {author} {\bibfnamefont {F.}~\bibnamefont {Nichele}},\ }\bibfield
  {title} {\enquote {\bibinfo {title} {Zero-energy modes from coalescing
  andreev states in a two-dimensional semiconductor-superconductor hybrid
  platform},}\ }\href {\doibase 10.1103/PhysRevLett.119.176805} {\bibfield
  {journal} {\bibinfo  {journal} {Phys. Rev. Lett.}\ }\textbf {\bibinfo
  {volume} {119}},\ \bibinfo {pages} {176805} (\bibinfo {year}
  {2017})}\BibitemShut {NoStop}%
\bibitem [{\citenamefont {Pientka}\ \emph {et~al.}(2017)\citenamefont
  {Pientka}, \citenamefont {Keselman}, \citenamefont {Berg}, \citenamefont
  {Yacoby}, \citenamefont {Stern},\ and\ \citenamefont
  {Halperin}}]{Pientka2017topological}%
  \BibitemOpen
  \bibfield  {author} {\bibinfo {author} {\bibfnamefont {Falko}\ \bibnamefont
  {Pientka}}, \bibinfo {author} {\bibfnamefont {Anna}\ \bibnamefont
  {Keselman}}, \bibinfo {author} {\bibfnamefont {Erez}\ \bibnamefont {Berg}},
  \bibinfo {author} {\bibfnamefont {Amir}\ \bibnamefont {Yacoby}}, \bibinfo
  {author} {\bibfnamefont {Ady}\ \bibnamefont {Stern}}, \ and\ \bibinfo
  {author} {\bibfnamefont {Bertrand~I.}\ \bibnamefont {Halperin}},\ }\bibfield
  {title} {\enquote {\bibinfo {title} {Topological superconductivity in a
  planar josephson junction},}\ }\href {\doibase 10.1103/PhysRevX.7.021032}
  {\bibfield  {journal} {\bibinfo  {journal} {Phys. Rev. X}\ }\textbf {\bibinfo
  {volume} {7}},\ \bibinfo {pages} {021032} (\bibinfo {year}
  {2017})}\BibitemShut {NoStop}%
\bibitem [{\citenamefont {Hell}\ \emph
  {et~al.}(2017{\natexlab{a}})\citenamefont {Hell}, \citenamefont {Leijnse},\
  and\ \citenamefont {Flensberg}}]{Hell2017two}%
  \BibitemOpen
  \bibfield  {author} {\bibinfo {author} {\bibfnamefont {Michael}\ \bibnamefont
  {Hell}}, \bibinfo {author} {\bibfnamefont {Martin}\ \bibnamefont {Leijnse}},
  \ and\ \bibinfo {author} {\bibfnamefont {Karsten}\ \bibnamefont
  {Flensberg}},\ }\bibfield  {title} {\enquote {\bibinfo {title}
  {Two-dimensional platform for networks of majorana bound states},}\ }\href
  {\doibase 10.1103/PhysRevLett.118.107701} {\bibfield  {journal} {\bibinfo
  {journal} {Phys. Rev. Lett.}\ }\textbf {\bibinfo {volume} {118}},\ \bibinfo
  {pages} {107701} (\bibinfo {year} {2017}{\natexlab{a}})}\BibitemShut
  {NoStop}%
\bibitem [{\citenamefont {Hell}\ \emph
  {et~al.}(2017{\natexlab{b}})\citenamefont {Hell}, \citenamefont {Flensberg},\
  and\ \citenamefont {Leijnse}}]{Hell2017Coupling}%
  \BibitemOpen
  \bibfield  {author} {\bibinfo {author} {\bibfnamefont {Michael}\ \bibnamefont
  {Hell}}, \bibinfo {author} {\bibfnamefont {Karsten}\ \bibnamefont
  {Flensberg}}, \ and\ \bibinfo {author} {\bibfnamefont {Martin}\ \bibnamefont
  {Leijnse}},\ }\bibfield  {title} {\enquote {\bibinfo {title} {Coupling and
  braiding majorana bound states in networks defined in two-dimensional
  electron gases with proximity-induced superconductivity},}\ }\href {\doibase
  10.1103/PhysRevB.96.035444} {\bibfield  {journal} {\bibinfo  {journal} {Phys.
  Rev. B}\ }\textbf {\bibinfo {volume} {96}},\ \bibinfo {pages} {035444}
  (\bibinfo {year} {2017}{\natexlab{b}})}\BibitemShut {NoStop}%
\bibitem [{\citenamefont {Haim}\ and\ \citenamefont
  {Stern}(2018)}]{Haim2018double}%
  \BibitemOpen
  \bibfield  {author} {\bibinfo {author} {\bibfnamefont {Arbel}\ \bibnamefont
  {Haim}}\ and\ \bibinfo {author} {\bibfnamefont {Ady}\ \bibnamefont {Stern}},\
  }\bibfield  {title} {\enquote {\bibinfo {title} {The double-edge sword of
  disorder in multichannel topological superconductors},}\ }\href@noop {}
  {\bibfield  {journal} {\bibinfo  {journal} {arXiv preprint arXiv:1808.07886}\
  } (\bibinfo {year} {2018})}\BibitemShut {NoStop}%
\bibitem [{\citenamefont {Hart}\ \emph {et~al.}(2017)\citenamefont {Hart},
  \citenamefont {Ren}, \citenamefont {Kosowsky}, \citenamefont {Ben-Shach},
  \citenamefont {Leubner}, \citenamefont {Br{\"u}ne}, \citenamefont {Buhmann},
  \citenamefont {Molenkamp}, \citenamefont {Halperin},\ and\ \citenamefont
  {Yacoby}}]{Hart2017controlled}%
  \BibitemOpen
  \bibfield  {author} {\bibinfo {author} {\bibfnamefont {Sean}\ \bibnamefont
  {Hart}}, \bibinfo {author} {\bibfnamefont {Hechen}\ \bibnamefont {Ren}},
  \bibinfo {author} {\bibfnamefont {Michael}\ \bibnamefont {Kosowsky}},
  \bibinfo {author} {\bibfnamefont {Gilad}\ \bibnamefont {Ben-Shach}}, \bibinfo
  {author} {\bibfnamefont {Philipp}\ \bibnamefont {Leubner}}, \bibinfo {author}
  {\bibfnamefont {Christoph}\ \bibnamefont {Br{\"u}ne}}, \bibinfo {author}
  {\bibfnamefont {Hartmut}\ \bibnamefont {Buhmann}}, \bibinfo {author}
  {\bibfnamefont {Laurens~W}\ \bibnamefont {Molenkamp}}, \bibinfo {author}
  {\bibfnamefont {Bertrand~I}\ \bibnamefont {Halperin}}, \ and\ \bibinfo
  {author} {\bibfnamefont {Amir}\ \bibnamefont {Yacoby}},\ }\bibfield  {title}
  {\enquote {\bibinfo {title} {Controlled finite momentum pairing and spatially
  varying order parameter in proximitized hgte quantum wells},}\ }\href@noop {}
  {\bibfield  {journal} {\bibinfo  {journal} {Nature Physics}\ }\textbf
  {\bibinfo {volume} {13}},\ \bibinfo {pages} {87} (\bibinfo {year}
  {2017})}\BibitemShut {NoStop}%
\bibitem [{\citenamefont {Ren}\ \emph {et~al.}(2018)\citenamefont {Ren},
  \citenamefont {Pientka}, \citenamefont {Hart}, \citenamefont {Pierce},
  \citenamefont {Kosowsky}, \citenamefont {Lunczer}, \citenamefont {Schlereth},
  \citenamefont {Scharf}, \citenamefont {Hankiewicz}, \citenamefont {Molenkamp}
  \emph {et~al.}}]{Ren2018topological}%
  \BibitemOpen
  \bibfield  {author} {\bibinfo {author} {\bibfnamefont {Hechen}\ \bibnamefont
  {Ren}}, \bibinfo {author} {\bibfnamefont {Falko}\ \bibnamefont {Pientka}},
  \bibinfo {author} {\bibfnamefont {Sean}\ \bibnamefont {Hart}}, \bibinfo
  {author} {\bibfnamefont {Andrew}\ \bibnamefont {Pierce}}, \bibinfo {author}
  {\bibfnamefont {Michael}\ \bibnamefont {Kosowsky}}, \bibinfo {author}
  {\bibfnamefont {Lukas}\ \bibnamefont {Lunczer}}, \bibinfo {author}
  {\bibfnamefont {Raimund}\ \bibnamefont {Schlereth}}, \bibinfo {author}
  {\bibfnamefont {Benedikt}\ \bibnamefont {Scharf}}, \bibinfo {author}
  {\bibfnamefont {Ewelina~M}\ \bibnamefont {Hankiewicz}}, \bibinfo {author}
  {\bibfnamefont {Laurens~W}\ \bibnamefont {Molenkamp}},  \emph {et~al.},\
  }\bibfield  {title} {\enquote {\bibinfo {title} {Topological
  superconductivity in a phase-controlled josephson junction},}\ }\href@noop {}
  {\bibfield  {journal} {\bibinfo  {journal} {arXiv preprint arXiv:1809.03076}\
  } (\bibinfo {year} {2018})}\BibitemShut {NoStop}%
\bibitem [{\citenamefont {Fornieri}\ \emph {et~al.}(2018)\citenamefont
  {Fornieri}, \citenamefont {Whiticar}, \citenamefont {Wenming}, \citenamefont
  {Mar{\'\i}n}, \citenamefont {Drachmann}, \citenamefont {Keselman},
  \citenamefont {Gronin}, \citenamefont {Thomas}, \citenamefont {Wang},
  \citenamefont {Kallaher} \emph {et~al.}}]{Fornieri2018evidence}%
  \BibitemOpen
  \bibfield  {author} {\bibinfo {author} {\bibfnamefont {Antonio}\ \bibnamefont
  {Fornieri}}, \bibinfo {author} {\bibfnamefont {Alexander~M}\ \bibnamefont
  {Whiticar}}, \bibinfo {author} {\bibfnamefont {Setiawan}\ \bibnamefont
  {Wenming}}, \bibinfo {author} {\bibfnamefont {El{\'\i}as~Portol{\'e}s}\
  \bibnamefont {Mar{\'\i}n}}, \bibinfo {author} {\bibfnamefont {Asbj{\o}rn~CC}\
  \bibnamefont {Drachmann}}, \bibinfo {author} {\bibfnamefont {Anna}\
  \bibnamefont {Keselman}}, \bibinfo {author} {\bibfnamefont {Sergei}\
  \bibnamefont {Gronin}}, \bibinfo {author} {\bibfnamefont {Candice}\
  \bibnamefont {Thomas}}, \bibinfo {author} {\bibfnamefont {Tian}\ \bibnamefont
  {Wang}}, \bibinfo {author} {\bibfnamefont {Ray}\ \bibnamefont {Kallaher}},
  \emph {et~al.},\ }\bibfield  {title} {\enquote {\bibinfo {title} {Evidence of
  topological superconductivity in planar josephson junctions},}\ }\href@noop
  {} {\bibfield  {journal} {\bibinfo  {journal} {arXiv preprint
  arXiv:1809.03037}\ } (\bibinfo {year} {2018})}\BibitemShut {NoStop}%
\bibitem [{\citenamefont {Tinkham}(1975)}]{Tinkham2004introduction}%
  \BibitemOpen
  \bibfield  {author} {\bibinfo {author} {\bibfnamefont {Michael}\ \bibnamefont
  {Tinkham}},\ }\href@noop {} {\emph {\bibinfo {title} {Introduction to
  superconductivity}}}\ (\bibinfo  {publisher} {New York: McGraw-Hill},\
  \bibinfo {year} {1975})\BibitemShut {NoStop}%
\bibitem [{\citenamefont {Alicea}\ \emph {et~al.}(2011)\citenamefont {Alicea},
  \citenamefont {Oreg}, \citenamefont {Refael}, \citenamefont {von Oppen},\
  and\ \citenamefont {Fisher}}]{Alicea2011non}%
  \BibitemOpen
  \bibfield  {author} {\bibinfo {author} {\bibfnamefont {Jason}\ \bibnamefont
  {Alicea}}, \bibinfo {author} {\bibfnamefont {Yuval}\ \bibnamefont {Oreg}},
  \bibinfo {author} {\bibfnamefont {Gil}\ \bibnamefont {Refael}}, \bibinfo
  {author} {\bibfnamefont {Felix}\ \bibnamefont {von Oppen}}, \ and\ \bibinfo
  {author} {\bibfnamefont {Matthew~PA}\ \bibnamefont {Fisher}},\ }\bibfield
  {title} {\enquote {\bibinfo {title} {Non-abelian statistics and topological
  quantum information processing in 1d wire networks},}\ }\href
  {http://www.nature.com/nphys/journal/v7/n5/abs/nphys1915.html} {\bibfield
  {journal} {\bibinfo  {journal} {Nat. Phys.}\ }\textbf {\bibinfo {volume}
  {7}},\ \bibinfo {pages} {412--417} (\bibinfo {year} {2011})}\BibitemShut
  {NoStop}%
\bibitem [{\citenamefont {Clarke}\ \emph {et~al.}(2011)\citenamefont {Clarke},
  \citenamefont {Sau},\ and\ \citenamefont {Tewari}}]{Clarke2011majorana}%
  \BibitemOpen
  \bibfield  {author} {\bibinfo {author} {\bibfnamefont {David~J.}\
  \bibnamefont {Clarke}}, \bibinfo {author} {\bibfnamefont {Jay~D.}\
  \bibnamefont {Sau}}, \ and\ \bibinfo {author} {\bibfnamefont {Sumanta}\
  \bibnamefont {Tewari}},\ }\bibfield  {title} {\enquote {\bibinfo {title}
  {Majorana fermion exchange in quasi-one-dimensional networks},}\ }\href
  {\doibase 10.1103/PhysRevB.84.035120} {\bibfield  {journal} {\bibinfo
  {journal} {Phys. Rev. B}\ }\textbf {\bibinfo {volume} {84}},\ \bibinfo
  {pages} {035120} (\bibinfo {year} {2011})}\BibitemShut {NoStop}%
\bibitem [{\citenamefont {Plugge}\ \emph {et~al.}(2017)\citenamefont {Plugge},
  \citenamefont {Rasmussen}, \citenamefont {Egger},\ and\ \citenamefont
  {Flensberg}}]{Plugge2017majorana}%
  \BibitemOpen
  \bibfield  {author} {\bibinfo {author} {\bibfnamefont {Stephan}\ \bibnamefont
  {Plugge}}, \bibinfo {author} {\bibfnamefont {Asbj{\o}rn}\ \bibnamefont
  {Rasmussen}}, \bibinfo {author} {\bibfnamefont {Reinhold}\ \bibnamefont
  {Egger}}, \ and\ \bibinfo {author} {\bibfnamefont {Karsten}\ \bibnamefont
  {Flensberg}},\ }\bibfield  {title} {\enquote {\bibinfo {title} {Majorana box
  qubits},}\ }\href@noop {} {\bibfield  {journal} {\bibinfo  {journal} {New
  Journal of Physics}\ }\textbf {\bibinfo {volume} {19}},\ \bibinfo {pages}
  {012001} (\bibinfo {year} {2017})}\BibitemShut {NoStop}%
\bibitem [{\citenamefont {Karzig}\ \emph {et~al.}(2017)\citenamefont {Karzig},
  \citenamefont {Knapp}, \citenamefont {Lutchyn}, \citenamefont {Bonderson},
  \citenamefont {Hastings}, \citenamefont {Nayak}, \citenamefont {Alicea},
  \citenamefont {Flensberg}, \citenamefont {Plugge}, \citenamefont {Oreg},
  \citenamefont {Marcus},\ and\ \citenamefont {Freedman}}]{Karzig2017scalable}%
  \BibitemOpen
  \bibfield  {author} {\bibinfo {author} {\bibfnamefont {Torsten}\ \bibnamefont
  {Karzig}}, \bibinfo {author} {\bibfnamefont {Christina}\ \bibnamefont
  {Knapp}}, \bibinfo {author} {\bibfnamefont {Roman~M.}\ \bibnamefont
  {Lutchyn}}, \bibinfo {author} {\bibfnamefont {Parsa}\ \bibnamefont
  {Bonderson}}, \bibinfo {author} {\bibfnamefont {Matthew~B.}\ \bibnamefont
  {Hastings}}, \bibinfo {author} {\bibfnamefont {Chetan}\ \bibnamefont
  {Nayak}}, \bibinfo {author} {\bibfnamefont {Jason}\ \bibnamefont {Alicea}},
  \bibinfo {author} {\bibfnamefont {Karsten}\ \bibnamefont {Flensberg}},
  \bibinfo {author} {\bibfnamefont {Stephan}\ \bibnamefont {Plugge}}, \bibinfo
  {author} {\bibfnamefont {Yuval}\ \bibnamefont {Oreg}}, \bibinfo {author}
  {\bibfnamefont {Charles~M.}\ \bibnamefont {Marcus}}, \ and\ \bibinfo {author}
  {\bibfnamefont {Michael~H.}\ \bibnamefont {Freedman}},\ }\bibfield  {title}
  {\enquote {\bibinfo {title} {Scalable designs for
  quasiparticle-poisoning-protected topological quantum computation with
  majorana zero modes},}\ }\href {\doibase 10.1103/PhysRevB.95.235305}
  {\bibfield  {journal} {\bibinfo  {journal} {Phys. Rev. B}\ }\textbf {\bibinfo
  {volume} {95}},\ \bibinfo {pages} {235305} (\bibinfo {year}
  {2017})}\BibitemShut {NoStop}%
\bibitem [{you()}]{youtube}%
  \BibitemOpen
  \href@noop {} {}\bibinfo {note} {The animation is also available on YouTube:
  \url{https://www.youtube.com/watch?v=itb4gRoE2H4&feature=youtu.be}.}\BibitemShut
  {Stop}%
\bibitem [{\citenamefont {Ivanov}(2001)}]{Ivanov2001non}%
  \BibitemOpen
  \bibfield  {author} {\bibinfo {author} {\bibfnamefont {D.~A.}\ \bibnamefont
  {Ivanov}},\ }\bibfield  {title} {\enquote {\bibinfo {title} {Non-abelian
  statistics of half-quantum vortices in $\mathit{p}$-wave superconductors},}\
  }\href {\doibase 10.1103/PhysRevLett.86.268} {\bibfield  {journal} {\bibinfo
  {journal} {Phys. Rev. Lett.}\ }\textbf {\bibinfo {volume} {86}},\ \bibinfo
  {pages} {268--271} (\bibinfo {year} {2001})}\BibitemShut {NoStop}%
\end{thebibliography}%
